\documentclass[manuscript]{aastex}
\usepackage{tikz}
\usepackage{amsmath}

\usepackage{graphicx}

\usepackage{pdflscape}

\usepackage{hyperref}

\usepackage{rotating}

\slugcomment{\textbf{The Astronomical Journal} }

\shorttitle{12P/Brooks}
\shortauthors{Jewitt}

\begin{document}

\title{Multiple Outbursts of Halley-Type Comet 12P/Pons-Brooks}

\author{
David Jewitt$^{1}$ and Jane Luu$^2$,
} 
\affil{$^1$Department of Earth, Planetary and Space Sciences, UCLA, Los Angeles, CA 90095-1567}
\affil{$^2$ Centre for Planetary Habitability (PHAB), Dept. of Geosciences, University of Oslo, NO-0315 Oslo, Norway}

\email{djewitt@gmail.com}

\begin{abstract}
We present optical observations of Halley type comet 12P/Pons-Brooks on its approach to perihelion.  The comet was active even in the first observations at $\sim$8 au. Starting at $\sim$4 au, 12P exhibited an extraordinary series of outbursts, in which the brightness changed by a factor up to 100 and the coma morphology transformed under the action of radiation pressure into a distinctive ``horned'' appearance.  Individual outburst dust masses are several $\times10^9$ kg, with  kinetic energies $\sim$10$^{14}$ J, release times $\sim$10$^4$ s and effective power $\sim$10$^{10}$ W.  These properties are most consistent with, although do not definitively establish, an origin by the crystallization of amorphous water ice with the related release of trapped supervolatile gases.  This interpretation is supported by the observation that the specific outburst energy and the specific crystallization energy are comparable (both  $\sim$10$^5$ J kg$^{-1}$).  
\end{abstract}


\section{INTRODUCTION}
\label{intro}
Comet 12P/Pons-Brooks (hereafter ``12P'') was discovered by Jean Louis Pons on 1812 July 21 and rediscovered by William Brooks one orbit later, on 1883 September 2 \citep{Kro03}.  The orbital elements include semimajor axis $a$ = 17.184 au, eccentricity $e$ = 0.955 and inclination $i$ = 74\degr.2, giving a Tisserand parameter with respect to Jupiter $T_J$ = 0.598, and an orbital period $P\sim$ 71 years. The orbit is thus distinct from those of the short-period comets (which have 2 $\le T_J \le$ 3) and from the long-period comets which have $T_J <$ 2 and $P >$ 200 years \citep{Don15}.  

12P is a ``Halley-type'' comet (HTC), a dynamically intermediate population having orbit periods 20 $\le P \le$ 200 years and Tisserand parameters $T_J <$ 2. The currently known HTCs number $\sim$10$^2$ and represent a tiny fraction of the $\sim$4000 historically recorded comets. Their likely source  remains to be firmly elucidated, with suggested contributions from  the Oort cloud \citep{Eme13} and the scattered disk component of the Kuiper belt \citep{Fer16}. Because of their relative rarity and long orbital periods (leading to long intervals between near-perihelion periods of observability), the HTCs are perhaps the least-well characterized of the major comet groups, both dynamically and physically. 

The best-studied HTC is the archetypal object 1P/Halley, whose orbital semimajor axis, $a$ = 17.928 au, and eccentricity, $e$ = 0.968, are remarkably similar to those of 12P (although the orbit of 1P/Halley differs in being retrograde, with $i$ = 162\degr.2).  The perihelion distances are $q$ = 0.781 au in the case of 12P, and 0.575 au in the case of 1P/Halley, meaning that both deeply penetrate the terrestrial planet region.  Backwards integrations of the motion of 12P reveal linkage to comets C/1385 U1, C/1457 A1 and, possibly, even to an unnamed comet observed in 245 AD \citep{Mey20}, indicating dynamical stability on $\sim10^3$ year timescales.  The lifetime of 12P to ejection from the solar system is of order 10$^6$ years, during which time dramatic transformations of the orbit, even from prograde to retrograde, are possible \citep{Bai96}.

Comet 12P was recovered on UT 2020 June 10 at 11.89 au \citep{Ye20}, about 4 years before reaching perihelion at $q$ = 0.781 au, on UT 2024 April 21.1.  These authors reported a 12P nucleus radius $r_n = 17\pm$6 km (geometric albedo 0.04 assumed), corresponding to absolute magnitude $H$ = 13.4. However, 12P was already active at the time so that the photometrically derived radius is only an upper limit to the true value.

 Data from the previous orbit (the last perihelion occurred on 1954 May 23) pre-date the era of electronic detectors and provide observations  that are limited in quantity and quality compared to those possible using the instrumentation of today.   For this reason, we initiated systematic observations to examine the development of activity in 12P upon approach to perihelion.  To our surprise, these observations showed the emergence of a series of dramatic photometric outbursts, each of which gave a distinctive ``horned" appearance to the comet.  The purpose of the present paper is to describe and interpret the pre-perihelion observations of 12P.

\section{OBSERVATIONS}
\begin{figure}
\epsscale{0.99}
\plotone{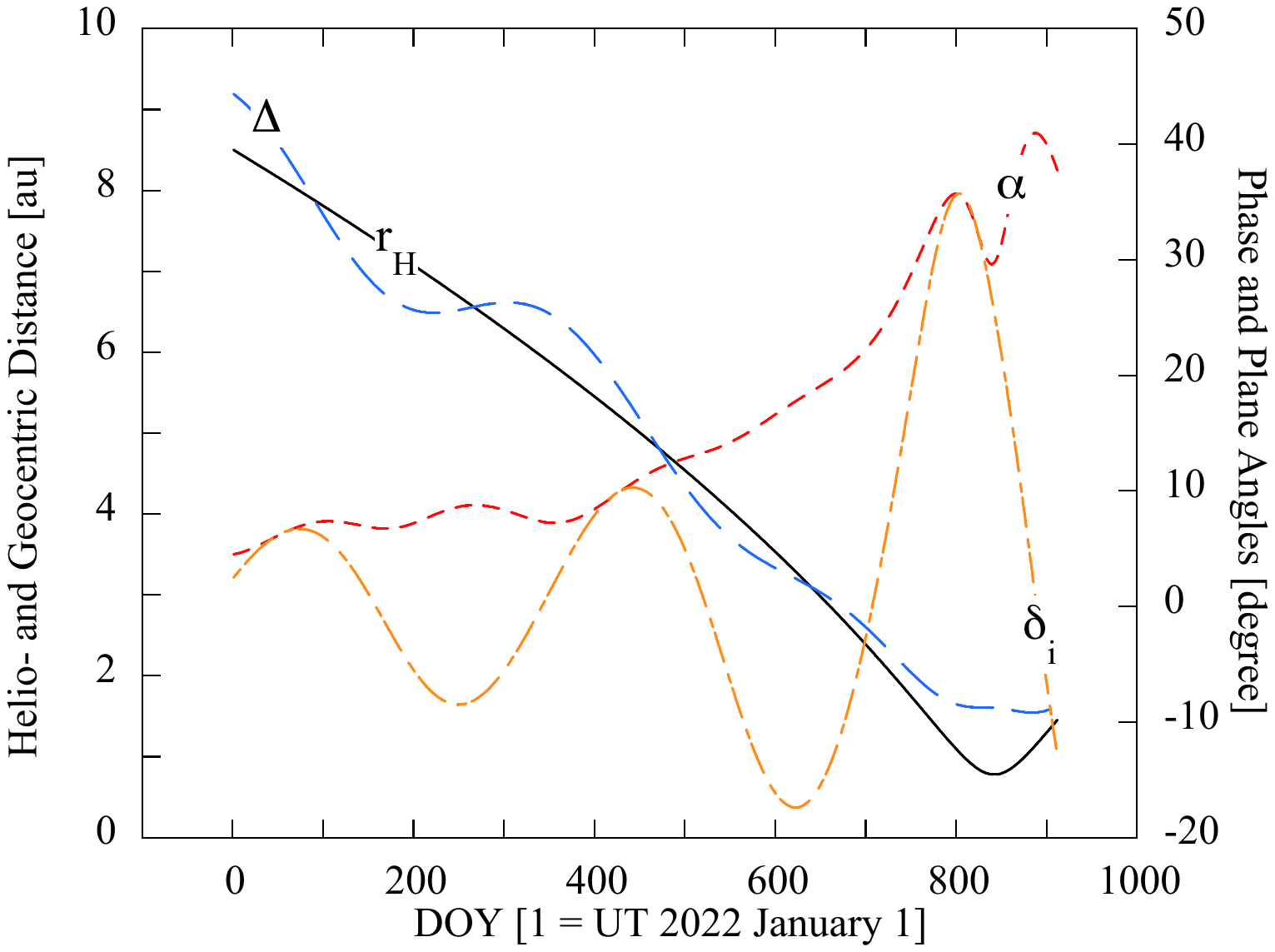}

\caption{Observing geometry as a function of Day of Year (1 = UT 2022 January 1).  Heliocentric ($r_H$) and geocentric distances ($\Delta$) shown as solid black and long-dashed blue lines, respectively, are referred to the left-hand vertical axis. The phase angle ($\alpha$) and orbital plane angle ($\delta_i$), shown as short-dashed red and dash-dot orange lines, respectively) refer to the right-hand axis.  \label{RDa}}
\end{figure}

We use data from two sources, with the common aim of examining the development of cometary activity in 12P upon its approach to the Sun.  Data from the first source, the Nordic Optical Telescope (NOT) are sparsely sampled but provide arcsecond resolution images across the heliocentric distance range from $r_H$ = 7.9 au inbound to 2.4 au.  The second source is a remarkable photometry compilation from the British Astronomical Association, which provides denser temporal sampling in the $\sim$4 au to $\sim$2 au range in the form of a series of fixed aperture photometric measurements.  The resolved and photometric time series data provide complementary views of the development of activity in 12P.  Figure \ref{RDa} shows the time-dependent geometry of the observations.

\subsection{Nordic Optical Telescope}
\begin{sidewaysfigure}
\epsscale{0.95}
\plotone{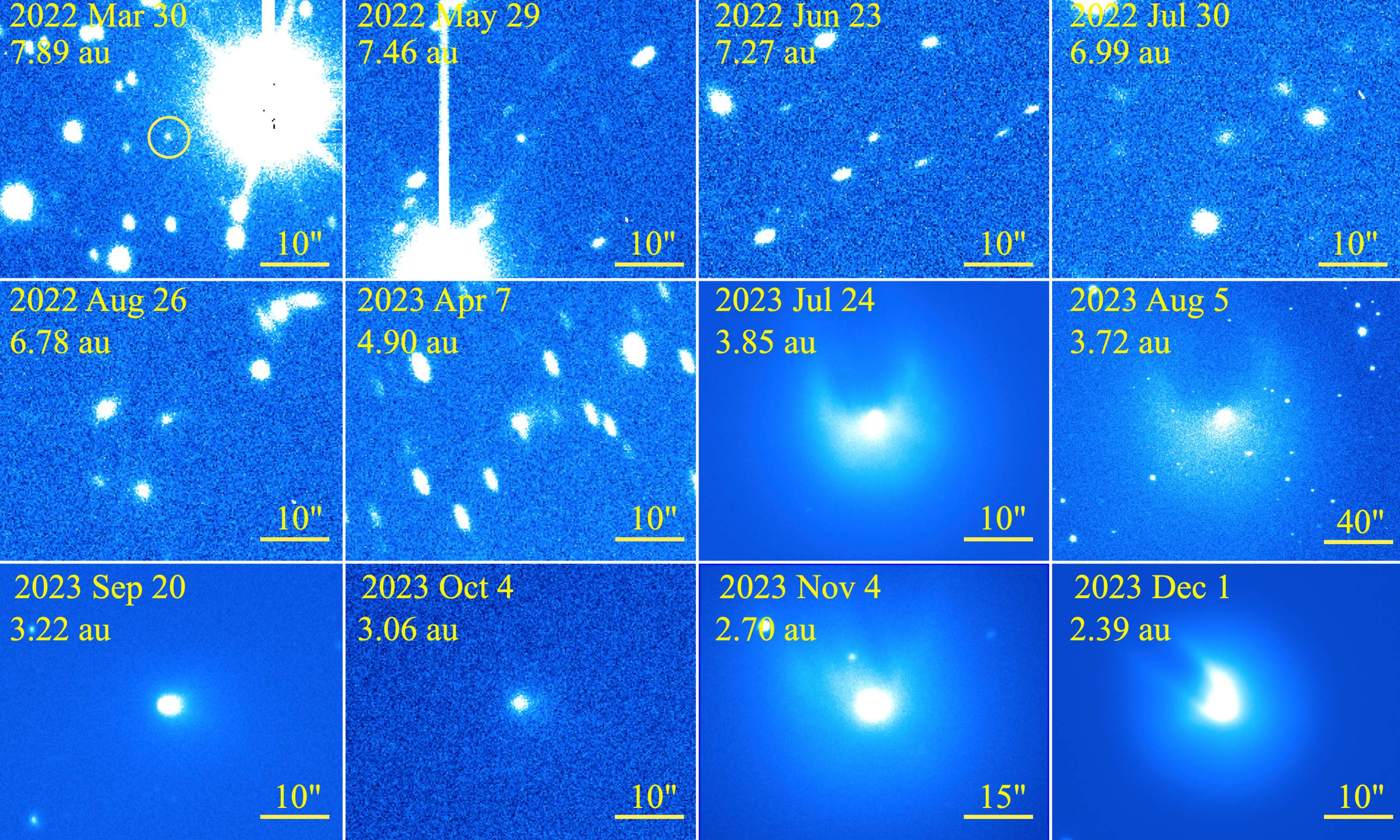}
\caption{Composite showing morphological development of 12P in a subset of the NOT R-band images, with the dates and heliocentric distances of each indicated.  The comet is centered in each panel. Note that different panels have different scales, as marked. \label{composite}}
\end{sidewaysfigure}

Observations were obtained using the 2.5m diameter NOT located on La Palma, in the Canary Islands.  We used the ALFOSC  (Andalucia Faint Object Spectrograph and Camera) imaging spectrometer camera at the $f$/11 Cassegrain focus, with the telescope tracked to follow the non-sidereal motion of the comet.  ALFOSC uses a 2048$\times$2064 pixel ``e2v Technologies'' charge-coupled device (CCD), yielding an image scale 0.214\arcsec~pixel$^{-1}$ across a 6.5\arcmin$\times$6.5\arcmin~field of view.  Seeing was variable from night to night but stayed mostly within the range 1.0\arcsec~to 1.5\arcsec~FWHM (full width at half maximum).
 We used a broadband  R filter ($\lambda_c$ = 6500\AA, $\Delta \lambda$ = 1300\AA) to measure the magnitude of 12P, with a majority of the data using integration times of 90s to 150 s.  Observations in this filter are relatively immune to contamination by resonance fluorescence from cometary gases and so provide a reliable measure of scattering from dust. Flat fields for these filters were constructed from images of an illuminated spot inside the dome of the NOT, while the bias level of the CCD was measured from a set of images obtained in darkness at the end of each night.  Photometric calibration was obtained from observations of Landolt stars \citep{Lan92}  and of field stars, using calibrations from the Gaia and Sloan DR14 digital sky surveys \citep{Pan22}.  For easy comparison with the BAA photometry, we used a NOT photometry aperture of 9\arcsec~angular radius where possible.  However, the comet was faint in the earlier observations  ($r_H \gtrsim$ 4 au), forcing the use of smaller photometry apertures to eliminate field objects and minimize the sky noise.  These smaller apertures (typically 2.1\arcsec~in radius with sky subtraction from a contiguous annulus 21.4\arcsec~wide), by excluding some of the extended emission from the comet, set lower limits to the brightness and scattering cross-section.  A journal of NOT observations is given in Table \ref{geometry} and a composite of selected images is shown in Figure \ref{composite}.
 
\subsection{British Astronomical Association Archive}
We additionally use a  compilation of magnitudes generously provided by Nick James on behalf of the British Astronomical Association\footnote{\url{https://britastro.org/section_news_item/12p-pons-brooks-latest-lightcurve}}. These magnitudes were obtained using CCDs on a variety of small telescopes, mostly without filters.  They have been calibrated using Gaia DR2 to effective Gaia Broadband (G) magnitudes.  The Gaia G band includes essentially the entire optical spectrum, with a center wavelength near 6000\AA~and full width at half maximum $\sim$4000\AA~\citep{Eva18}.  The conversion between Gaia G and broadband R for a solar-type spectrum is $R \sim G - 0.2$ \citep{Eva18}; the difference is so small given the range of observed magnitudes that we have ignored it for our present purposes.  The BAA measurements are standardized to a fixed aperture 9\arcsec~in radius.  This removes one of the main sources of ambiguity in the interpretation of otherwise heterogeneous photometry. However,the interpretation remains susceptible to uncertainty caused by the extended nature of the comet and its (time variable) radial surface brightness gradient, as we discuss below.  The NOT and BAA measurements are plotted as a function of date in Figure \ref{NOT_BAA_4}; the NOT data have upward pointing vertical bars to indicate that the measurements are lower limits forced by the use of small apertures.

\section{DISCUSSION}

\textbf{Morphology:}  Figure \ref{composite} is a composite of selected NOT images showing the morphological development of 12P.  The comet is centrally placed in each panel and the scale has been adjusted to account for the varying angular size of the coma, as indicated by the yellow scale bars.  12P was marginally resolved in observations from 7.89 au (2022 March 30) and 7.46 au (2022 May 29) but had developed a clear radiation pressure swept tail by 6.99 au (2022 Jul 30).  The morphology changed dramatically on 2023 July 24 (3.85 au) in conjunction with an outburst, the coma developing a ``horned" shape.  The ``horned" appearance faded by 2023 September 20, but reappeared in subsequent observations, ending 2023 December 1 (2.39 au).

We defer a detailed study of the time-dependent coma morphology to a later paper and confine ourselves here to several important coma characteristics that are revealed by a simple inspection of the data.  The  morphology of outburst near 2023 July 24 (hereafter called Outburst A) is shown on six dates spanning 24 days in Figure \ref{horns}.  The images show an expanding, upside-down umbrella-like coma, formed by dust ejected from the hot, Sun-facing side of the nucleus and then accelerated anti-sunward by radiation pressure.  The ``horned'' appearance of the comet results from projection into the plane of the sky of this dust as it is pushed back, and gives the appearance of a dark or hollow zone behind the nucleus.  Close inspection of the images -- particularly 2023 July 28, August 5 and 9 -- reveals that the umbrella is spatially structured, indicating inhomogeneities in the ejected material presumably related to structure at the coma source.

We used the images in Figure \ref{horns} to measure the expansion of the coma along its symmetry axis, using 2023 July 20 as the baseline.  This measurement is intrinsically difficult because the expanding coma is not sharp-edged; we estimate that the uncertainties are $\sim$10\%.  The measurements, shown in Figure \ref{nose_length}, are consistent with constant velocity expansion. A least-squares fit to the slope gives sky-plane speed $V_{s}$ = 97$\pm$10 m s$^{-1}$, which is a lower limit to the true speed of the coma particles because of the effects of projection.  An accurate deprojection would require  the use of a 3-dimensional coma model, with intrinsic uncertainties remaining due to the unknown angular velocity distribution of the ejected material.  A crude estimate can be obtained by assuming that the coma axis of symmetry is parallel to the Sun-comet line. Then the expansion speed projected into the sky plane is related to the true speed by $V = V_{s}/\sin{\alpha}$, where $\alpha$ is the phase angle.  With $\alpha = 15\degr$ at the time of the observations in Figure \ref{horns}, we infer $V$ =  375$\pm$38 m s$^{-1}$.  

While $V_s$ is measured to an accuracy of about $\pm$10\%, the true uncertainty on the deprojected velocity, $V$, is difficult to define.  \cite{Bob32} summarized measurements of expanding ``halos'' in 12P from the 1884 perihelion (i.e., two orbits before the present one).  The best of these, taken by Otto Struve when 12P was at $r_H$ = 2.12 au and at phase angle $\alpha$ = 27\degr, give sky-plane speed $V_s \sim$ 280 m s$^{-1}$. Using the same deprojection as for our own data, the inferred ejection speed is $V = V_s/\sin(\alpha) \sim$ 620 m s$^{-1}$.   All else being equal, the terminal speed of ejected particles should scale with heliocentric distance as $V \propto r_H^{-1}$ (see derivation in Section 3 of \cite{Jew21b}).  Therefore, scaling from 2.12 au to  3.4 au, the Struve measurement gives $V \sim$ 390 m s$^{-1}$, to be compared with $V$ =  375$\pm$38 m s$^{-1}$  deduced from our own data.  The agreement obtained by different observers at different heliocentric distances and phase angles is reassuring and, indeed, probably better than should be expected. We conclude only that there is no evidence from this comparison that our estimate of the dust expansion speed, $V$, is grossly incorrect.

The lack of appreciable deceleration of the dust by solar radiation pressure allows us to set a lower limit to the effective dust particle size.  Neglecting the orbital motion of the cometary nucleus, the distance traveled by a dust particle launched towards the Sun at speed $V$ and pushed away by radiation pressure is simply described by 

\begin{equation}
\ell = V \delta t - \frac{\beta g_{\odot}(r_H) \delta t^2}{2}
\label{ell}
\end{equation}

\noindent where $\delta t = t - t_0$ is the time elapsed since the moment of ejection, $t_0$.  Quantity $\beta$ is the ratio of the acceleration due to radiation pressure to the acceleration due to solar gravity, $g_{\odot}(r_H)$.  The particle radius, $a$,  and $\beta$ are approximately related by $\beta \sim 10^{-6}/a$, where $a$ is expressed in meters.  We take $g_{\odot}(3.7~\textrm{au}) = 4.3\times10^{-4}$ m s$^{-2}$ and plot Equation \ref{ell} for $\beta = 10^{-3}, 10^{-2}, 10^{-1}$ and 1 in Figure \ref{nose_length}.  Values $\beta \sim$ 1 ($a \sim$ 1 $\mu$m) are ruled out by the linearity of the coma expansion and we conclude that acceptable fits have $\beta \lesssim$ 0.1, consistent with particle radii $a \gtrsim$ 10 $\mu$m.  Smaller grains (larger $\beta$) are no doubt present but in numbers insufficient to define the leading edge of the expanding coma.  Their apparent absence may reflect, in part, the presence of inter-particle cohesion, which preferentially binds small particles.  Small grains also travel faster than large grains, under the action of gas drag, and so leave the coma on a shorter timescale.  Larger grains (smaller $\beta$) simply add to the coma surface brightness inside the coma.

For comparison with $V$, the average thermal speed of gas molecules sublimated from the nucleus is given by $V_{th} = (8 k T/(\pi \mu m_H)^{1/2})$, where $k = 1.38\times10^{-23}$ J K$^{-1}$ is Boltzmann's constant, $T$ is the local sublimation temperature, $\mu$ is the molecular weight of the gas molecules and $m_H = 1.67\times10^{-27}$ kg is the mass of the hydrogen atom.  Substituting $T$ = 168 K (calculated using Equation B1, Appendix B, for equilibrium sublimation of water ice from the sun-facing hemisphere of a spherical nucleus at 3.7 au), $\mu$ = 18 (for H$_2$O), gives $V_{th}$ = 440 m s$^{-1}$.  Evidently, $V$ and $V_{th}$ are comparable, indicating that the 10 $\mu$m particles ejected by 12P are small enough to be dynamically well-coupled to the outflowing gas. 

We assume that solid particles are expelled from 12P by gas drag forces, in which case the terminal velocity, $V(a)$, is related to the particle size, $a$, by

\begin{equation}
V(a) = V_0 \left(\frac{a_o}{a}\right)^{1/2}
\label{speed}
\end{equation}

\noindent where $V_0$ and $a_0$ are constants.  Based on the outburst measurements described above, we can set $V_0$ = 1.2 m s$^{-1}$ in Equation \ref{speed} such that  $V(a)$ = 375 m s$^{-1}$ when $a$ = 10$^{-5}$ m and $a_0$ = 1 m.

\textbf{Outburst Decay Shapes:}
The long term photometry of 12P shows a prominent series of outbursts (Figure \ref{BAA}), all of which have a characteristic sharp rise to maximum followed by a more protracted fading back towards the base brightness levels.  The major outbursts are labeled A - G in the Figure.   
The rise time for most outbursts is short compared to the (roughly) daily sampling of the BAA photometry, while the fading phase is resolved and can be followed for several tens of days. 

Table \ref{outbursts} summarizes the basic parameters of each outburst.  The date and DOY of the best estimate of the time of peak brightness were obtained by inspection of the photometry, as well as $V_{peak}$ and $V_{base}$, the peak magnitude and the magnitude at the base determined by interpolation.  The cross-sections of the coma at the peak and the base are calculated from Equation \ref{inversesquare}; we also list $\Delta C = C_{peak} - C_{base}$, the increase in cross-section caused by each outburst.

The cross section of particles inside the photometry aperture is

\begin{equation}
C(t) = \int_{a_{min}}^{a_{max}} \pi a^2 n(a)da
\label{C}
\end{equation}

\noindent where $n(a) da$ is the number of particles with radii in the range $a \rightarrow a+da$.  We write $n(a)da = \Gamma a^{-\gamma}da$, with $\Gamma$ and $\gamma$ being constants,  and consider particle radii in the range $a_{min} \le a \le a_{max}$. In gas drag acceleration, the smallest particles are the fastest (Equation \ref{speed}) and so $a_{min}$  increases with the time since ejection as smaller particles escape the projected photometry aperture.  The residence time of particles within the projected photometry aperture is 

\begin{equation}
t(a) = s \phi \Delta/V(a),
\label{tres}
\end{equation}

\noindent where $\phi$ is the photometry aperture radius in arcseconds, $s = 7.25\times10^5$ m (arcsec)$^{-1}$ is the distance subtending 1\arcsec~from 1 au and geocentric distance $\Delta$ is in au.  Substituting for $V(a)$ from Equation \ref{speed} we find that the smallest particles remaining in the photometry aperture at post-outburst time $\delta t$ have radius $a_{min}$ given by

\begin{equation}
a_{min} = a_0 \left(\frac{V_0 \delta t}{s \phi \Delta}\right)^2
\label{amin}
\end{equation}

\noindent Equation \ref{C} and \ref{amin} together give

\begin{equation}
C(t) =  \frac{\pi \Gamma a_0^2}{\gamma-3}  \left(\frac{V_0 \delta t}{\phi s \Delta}\right)^{6-2\gamma}
\label{Coft}
\end{equation}

\noindent provided $\gamma > 3$, $a_{max} \gg a_{min}$ and $\delta t >$ 0.  Equation \ref{Coft} shows that the scattering cross-section in an outburst should vary with time since ejection as $\delta t^{6-2\gamma}$; therefore, measurement of the fading can constrain the size distribution of the particles, subject to the assumptions that they are impulsively ejected, accelerated by gas drag and distributed by size as a power law.  We note that Equation \ref{Coft} has the same $t^{6-2\gamma}$ time dependence as derived for the fading of a rotationally disrupted asteroid, because  radiation pressure also induces a particle speed $\propto a^{-1/2}$ relation \citep{Jew17}. 

The best observed outburst is D, for which the rise and fall sections of the profile are relatively well characterized by a large number of measurements and confusion with nearby outbursts is minimal.  We enlarge the Outburst D profile in  Figure \ref{outburst_D}, while Figure \ref{Burst_D}  shows the same outburst but with the data plotted on two logarithmic axes with the underlying continuum gradient removed.  The dashed black line in Figure 8 shows a power law fitted by least squares, and weighted assuming 5\% photometric uncertainties.  The time axis has been shifted so that $\delta t$ = 0 corresponds to the time of the brightest measurement and the measured brightnesses have been normalized to unity at the peak.  For Outburst D, the time of peak brightness is relatively well constrained to lie within $\pm$0.1 day of $t_0$ = 685.0 (UT 2023 November 16.0). The best fit gives a time dependence $C(t) \propto t^{-2.5}$, corresponding to $\gamma = 4.2\pm0.2$ by Equation \ref{Coft}, where the uncertainty reflects the uncertainty on $t_0$.  The fit is not perfect but, given the quality of the data and the approximations of the model, adequately matches the observations.

\noindent Figure \ref{outburst_D} shows that the fading is measurable to at least 10 days (8.6$\times10^5$ s) post-outburst. Substituting into Equation \ref{amin} with $\Delta$ = 2.6 au shows that only particles with radii $a \gtrsim$ 4 mm  remain in the 9\arcsec~photometry aperture for longer than 10 days unless resupplied by continuing nucleus activity.  Therefore, we assume that  the particle size distribution  spans the range from $a_{min} \sim$ 10 $\mu$m  to $a_{max} \gtrsim$ 4 mm, by these arguments, which safely satisfies our assumption (in Equation \ref{Coft}) that $a_{max} \gg a_{min}$.  

\textbf{Dust Mass:} The  mass of a collection of spheres of density $\rho$ and presenting a cross-section, $C$, is given by

\begin{equation}
M = \frac{4}{3} \rho \overline{a} C
\label{mass}
\end{equation}

\noindent where $\overline{a}$ is the  mean particle radius.  The mean particle radius is weighted by the cross-sections of particles occupying a differential size distribution $n(a) da = \Gamma a^{-\gamma} da$ giving

\begin{equation}
\overline{a} = \frac{\int_{a_{min}}^{a_{max}}  a \Gamma \pi  a^2 a^{-\gamma}da}{\int_{a_{min}}^{a_{max}} \Gamma  \pi a^2 a^{-\gamma}da}.
\end{equation}

Substituting  $\gamma$ = 4.2 and assuming $a_{max} \gg a_{min}$, as above, this relation gives $\overline{a}$ = 11/6 $a_{min}$ which, with $a_{min}$ = 10 $\mu$m gives $\overline{a}$ = 18 $\mu$m.  Then, with $\rho = 10^3$ kg m$^{-3}$, Equation \ref{mass} reduces to

\begin{equation}
M \sim 0.02 C.
\label{mass2}
\end{equation}

\noindent Making use of Equation \ref{mass2}, and assuming that the seven outbursts in Table \ref{outbursts} have the same size distribution as Outburst D, we compute masses from the excess cross-sections, as listed in Table \ref{outbursts}.  The mean outburst cross-section is $\overline{\Delta C} = (1.3\pm0.4)\times 10^{11}$ m$^2$ 
(median value 1.1$\times10^{11}$ m$^2$) and  the mean outburst mass $M \sim 2.6\times10^9$ kg (median $2.2\times10^9$ kg).   

To set 12P's outbursts in context, we note that outbursts measured in 67P/Churyumov-Gerasimenko ($\le10^4$ kg; \cite{Lin17}) and 46P/Wirtanen (10$^4$ to 10$^6$ kg; \cite{Kel21}) were comparatively tiny. Indeed, the  mass of material released in a single outburst of 12P rivals the mass lost from 67P over the course of an \textit{entire} orbit (estimated as $\sim$3$\times10^9$ kg \citep{Ber15} to 5$\times10^9$ kg \citep{Dav22}).  The outbursts of 1P/Halley ($\sim10^9$ kg; \citep{Sek92}), 15P/Finlay (10$^8$ to 10$^9$ kg; \cite{Ish16}), P/2010 H2 (Vales) ($\sim1.2\times10^9$ kg; \cite{Jew20}) and 29P/Schwassmann-Wachmann 1 ($\ge1.8\times10^9$ kg; \cite{Sch17}) are more similar in mass to the mean value derived here for 12P.  At the other extreme, the extraordinary  UT 2007 October 24 outburst of 17P/Holmes  had a  mass  (20–900)$\times10^{9}$ kg \citep{Li11}.  
 
   \citet{Gri25} reported outburst masses for 12P in the range $10^{10}$ to $10^{13}$ kg.  These authors assumed a smaller size index ($\gamma$ = 3.5 vs.~4.2 measured here) and particle sizes $\sim$1 mm  (i.e., 100 times larger than measured) to infer very large outburst masses, exceeding by an order of magnitude even that from 17P/Holmes.  We note that millimeter sized particles would be difficult to accelerate by gas drag to the high measured expansion speed of the coma.

\textbf{Heliocentric Lightcurve:}
The composite lightcurve (Figure \ref{NOT_BAA_4}) covers the heliocentric distance range 1.5 $\le r_H \le$ 8 au and spans $\sim$14 magnitudes.  The BAA and NOT photometry, where they overlap in time, are in good agreement.  


The dashed black line in Figure \ref{NOT_BAA_4} shows the expected magnitude of a nucleus having  $H$ = 13.4 computed from Equation \ref{inversesquare}.  Evidently, all the measurements in the figure are brighter than the $H$ = 13.4 model, consistent with the presence of coma.  The difference between the comet brightness and the nucleus model increases from $\sim$1 magnitude in the early observations to $\sim$6 magnitudes near 1 au, as the outgassing activity grows.  Superimposed on the overall brightening are a series of outbursts, resulting from impulsive mass loss events.   

First we discuss the overall lightcurve, then we address the outbursts. An immediate conclusion from the figure is that the activity in 12P is not entirely controlled by the sublimation of water ice.  This must be the case because the comet is active at distances $r_H \gtrsim$5 au, where water is not volatile.  As in other comets, likely contributors to this distant activity include the sublimation of carbon monoxide (CO) and dioxide (CO$_2$) ices.  To model the lightcurve we assume that the scattering cross section varies as $C(r_H) = C(1) r_H^{-m}$, with $C(1)$ being the cross section at $r_H$ = 1 au and $m$ being the heliocentric index.  In practice, the outgassing rate is a complicated and unknowable function of many different physical processes operating on the nucleus, and there is no compelling reason to expect that a single power law for $C(r_H)$ should adequately represent the data.  Despite this, as shown in Figure \ref{NOT_BAA_4}, we find comparably good matches to the non-outburst photometry with Equation \ref{R_c} and $m,n$ = 4.0, 2.0 and $m,n$ = 4.5, 1.0.    These two models emphasize that the heliocentric index, $m$, and the surface brightness index, $n$, are interdependent, and give an idea of the likely uncertainty on these indices.   

To estimate the  mass loss rate between outbursts we use Equation \ref{mass2} to estimate the mass from the cross section, then we divide by the photometry aperture residence time for dust particles, given by Equation \ref{tres}.  We use the sky-plane speed $V_s$ = 97 m s$^{-1}$ as measured for the dominant outburst particles, finding the residence time $t \sim$ 3.8 day at $r_H$ = 4 au to 1/2 day at 1.8 au.  The resulting mass loss rates, $dM/dt$ [kg s$^{-1}$], are well fitted by a power law relation

\begin{equation}
\frac{dM}{dt} = (6\pm1)\times10^3 r_H^{-2.8.\pm0.2}
\label{dmbdt}
\end{equation}

\noindent as shown in Figure \ref{dMbdt_vs_rH}.  The index in Equation \ref{dmbdt} differs from $m$ because the residence time in a fixed angular aperture varies in proportion to geocentric distance, $\Delta$, and $\Delta$ is correlated with $r_H$ (c.f., Figure \ref{RDa}).


\textbf{Nucleus: }  As noted in the introduction, optical photometry by \cite{Ye20} sets a  limit to the absolute magnitude of the nucleus $H > 13.4$ (equivalently, an upper limit to the nucleus radius, $r_n \lesssim$ 17 km (albedo 0.04 assumed)).  This limit is not contradicted by any of the data presented in the current paper, as is shown by the dashed black line in Figure \ref{NOT_BAA_4}.

The momentum lost in the outbursts should result in recoil of the nucleus, potentially observable as a non-gravitational acceleration.  By conservation of momentum, the maximum recoil speed of the nucleus resulting from ejection of the  total outburst mass, $M_O$, at speed, $V$, is $\Delta V = (M_O/M_n) V$, where $M_n$ is the mass of the nucleus.  With outbursts spread  over an interval, $\Delta t$,  the time-averaged acceleration is $\alpha_{NGA} \sim \Delta V/\Delta t$. For a spherical nucleus of radius, $r_n$, and density, $\rho$, 

\begin{equation}
\alpha_{NGA} \sim \frac{3 M_O V}{4 \pi \rho r_n^3 \Delta t}  
\label{alphanga}
\end{equation}

\noindent We take total outburst mass $M_O = 1.8\times10^{10}$ kg (Table \ref{outbursts}), $V$ = 375 m s$^{-1}$, $\rho$ = 10$^3$ kg m$^{-3}$, $\Delta t \sim$ 200 days (seven outbursts between DOY 568 and 748, c.f., Table \ref{outbursts}), to find $\alpha_{NGA} \sim 10^{-7}/r_n^{3}$ [m s$^{-2}$], with $r_n$ expressed in kilometers.    
Unfortunately, no measurement of $\alpha_{NGA}$ exists for 12P. Observational prospects for the detection of $\alpha_{NGA}$ (and inference of $r_n$) are dim given that  problems in measuring the center of light in a bright, diffuse moving target with highly variable and asymmetric morphology will be difficult to overcome.

Measurements of short-period comets show that a fraction of the outflow momentum from sublimated material, $k_T \sim$ 0.007, is non-radial and exerts a torque on the nucleus causing a change in the spin \citep{Jew21}.  By analogy, the outbursts from 12P presumably exert a torque on the nucleus and, over time, will cause the nucleus to enter a rotationally excited (non-principal axis) state.  Time-resolved observations of the nucleus taken in the absence of coma (e.g., near aphelion distances $\sim33$ au) might show evidence for excited rotation, and would be additionally useful in better defining the nucleus radius.

\textbf{Outburst Mechanism:}
Numerous cometary outburst mechanisms (e.g., impacts, runaway exothermic chemical reactions, unstable nucleus landforms, exploding supervolatile pressure pockets, crystallization) have been proposed  but few are observationally constrained.  Particularly detailed, close-up observations of outbursts are available for the nucleus of 67P/Churyumov-Gerasimenko \citep{Lin17}. There, outbursts have been tentatively linked to explosions from CO$_2$ rich sub-surface ``cavities'', and to cliff collapse \citep{Mul24}.   In cliff collapse, lateral erosion on the nucleus produces cliffs (and overhangs) in material so weak that even nucleus gravity ($\sim10^{-4}$ m s$^{-2}$ on a 1 km nucleus) can trigger instability and collapse.  The resulting debris sheet exposes previously buried ices to the Sun, which then sublimate, driving a surge in gas production. In-situ observations from 67P might be expected to inform the outburst process in 12P.  However, in view of the small scale of the outbursts in 67P (outburst masses $\sim10^{5}$ times smaller than in 12P), it is not obvious that either process can account for the outbursts in 12P.  For example, it is not obvious why or how cliffs or other vertical structures on 12P would remain stable until reaching critical collapse masses $\sim10^5$ times greater than in 67P.

The crystallization of amorphous ice offers a more promising mechanism for outbursts in comets generally \citep{Pri24}, and in 12P specifically.  Rearrangement of the crystalline structure from the amorphous state leads to the expulsion of large quantities of trapped gas \citep{Gud23}, potentially building the subsurface pressure to the point of rupture. Crystallization is exothermic ($\sim10^5$ J kg$^{-1}$), and is delayed by the time needed for externally applied heat to reach the amorphous ice, presumably buried beneath the exposed nucleus surface.  Thermal runaway due to crystallization has already been proposed to account for a distant outburst in outbound 1P/Halley \citep{Pri92}. We briefly address observational constraints on the outburst mechanism in 12P.

First, as shown in Figure \ref{excess}, the excess brightness at the peak of each outburst (local background subtracted), $\Delta V$ [magnitudes], seems to vary systematically with heliocentric distance.   However, inspection of  Table \ref{outbursts} shows that the cross-section of each outburst is always $\Delta C \sim 10^{11}$ m$^2$, independent of $r_H$.  In other words, the sizes of the outbursts remain roughly constant but they change the brightness by a larger factor when the comet is more distant and less active, causing the trend seen in Figure \ref{excess}. 

Second, the similarity of the  outburst cross-sections ($\sim10^{11}$ m$^2$) and masses ($\sim10^9$ kg) of the seven 12P outbursts (Table \ref{outbursts}) may suggest the existence of a fundamental physical scale in the nucleus of this comet.  Outbursts smaller by an order of magnitude  would be readily apparent in the existing data (Figure \ref{outburst_D}), yet few are evident.  Therefore, the similarity of outburst cross-sections cannot be produced by observational bias against the detection of smaller outbursts.

Third, the spacings of the outbursts are non-randomly distributed, as may be seen in Figure \ref{BAA}.  The data hint that outbursts (e.g., the pairs CD, DE and EF) tend to be separated by $\sim$14 days, or small integer multiples thereof.  If this spacing is real it may suggest that 14 days is a recharge time for the outburst mechanism, perhaps related to the propagation of conducted heat from the surface to buried supervolatiles.  However, given the small number of outbursts, it would be premature to pursue this suspicion here.

The best observed outburst, D, offers more insight into the outburst mechanism.  With mass $M = 6.8\times10^9$ kg (Table \ref{outbursts}) and ejecta velocity $V \sim$ 375 m s$^{-1}$, the kinetic energy of the outburst material is $E \sim 4.8\times10^{14}$ J. The rise-time of the outburst (Figure \ref{outburst_D}) is $t \sim$ 3 hours (1.1$\times10^4$ s), giving an average outburst mass ejection rate $M/t \sim 6\times10^5$ kg s$^{-1}$ ($\sim$10$^5$  times values  measured in 67P/Churyumov-Gerasimenko; \cite{Lin17}). The averaged outburst power is $P = E/t \sim 4.3\times10^{10}$ W.  If sustained in equilibrium, this would correspond to the solar power absorbed over an area $\sim$ 200 km$^2$, which equals the full area of the Sun-facing hemisphere of a 5.6 km radius nucleus.  More likely, disequilibrium prevails, and the outburst energy has been stored in the near-surface material by a chemical or other process.  

The kinetic energy per unit mass in Outburst D is $\sim7\times10^4$ J kg$^{-1}$, very close to the $10^5$ J kg$^{-1}$  released by the crystallization of amorphous ice (see discussion in \cite{Pri24}).  This is also true for outbursts in comets 17P/Holmes \citep{Li11}, P/2010 H2 (Vales) \citep{Jew20}, 332P/Ikeya-Murakami \citep{Jew16} and 15P/Finlay \citep{Ish16}.  If it is not a coincidence, this observation is consistent with energy from crystallization driving the outburst.  The $\sim$14-day interval between outbursts might then reflect  the time needed for heat to conduct through a crystalline and/or refractory mantle to trigger crystallization of the amorphous ice beneath.  

Downward propagation of crystallization should be limited by the steep temperature gradient in the cometary nucleus.  The e-folding distance scale for conduction of heat deposited on the surface, $d$, is determined by the diffusivity, $\kappa$ [m$^2$ s$^{-1}$] and by the timescale over which the heat is applied, $\tau$, given by $d \sim (\kappa \tau)^{1/2}$.  For example, the $\sim$14-day interval between outbursts might reflect the time needed for heat to conduct through a crystalline and/or refractory mantle to trigger crystallization of the amorphous ice beneath.  With thermal diffusivity $\kappa$ = 10$^{-8}$ m$^2$ s$^{-1}$, appropriate for a highly porous dielectric material, and $\tau$ = 14-day (1.2$\times10^6$ s) we estimate $d \sim$ 0.1 m.  We equate the volume of the outburst ejecta, $M/\rho$, to the volume of a circular patch of radius, $L$, and thickness, $d$, to find

\begin{equation}
L = \left(\frac{M}{\pi \rho d}\right)^{1/2}.
\end{equation}

\noindent Substituting $M \sim 7\times 10^9$ kg for Outburst D, $\rho$ = 10$^3$ kg m$^{-3}$ and $d$ = 0.1 m gives $L \sim$ 5 km. The nucleus of 12P cannot be much smaller than this if crystallization is the cause of the outbursts.

The rate of crystallization of amorphous ice is a strong function of the ice temperature (see review in \cite{Pri24}).  Peak subsolar temperatures on 12P at perihelion  exceed 400 K, and crystallization there would be effectively instantaneous.    Nevertheless, the survival of primordial amorphous ice in the subsurface regions of 12P should not be a surprise.  The  low diffusivity of cometary material delays and impedes the transmission of high temperatures from the illuminated surface to the interior.  Indeed, the deep interior of the nucleus likely retains the temperature of the cometary source region, whether it be the scattered disk component of the Kuiper Belt ($T \sim$ 30 K to 40 K) or the Oort cloud ($T \sim$ 10 K); and for ice formed at these low temperatures,  amorphous structure is  thermodynamically preferred.  Progressive ablation of the surface by sublimation, amounting to several meters per orbit, can maintain amorphous ice  a few thermal skin depths below the surface.  

While these considerations are not enough to conclude that the outbursts of 12P are driven by crystallization, the available data are consistent with this possibility.  As noted in the introduction, 12P and 1P/Halley are (currently) dynamically similar but 12P is more prone to outbursts than 1P/Halley, perhaps because amorphous ice persists closer to its sublimating surface. The question of whether this difference is evolutionary or primordial remains unaddressed.

\clearpage

\section{SUMMARY}
We present pre-perihelion observations of the Halley type comet 12P/Pons-Brooks.  

\begin{itemize}
\item 12P was active over the full 8 au to 1 au heliocentric distance range, requiring that mass loss be due, at least in part, to the sublimation of ices more volatile than water ice.  The mass loss rate between outbursts followed an approximately $r_H^{-2.8\pm0.2}$ variation.

\item A series of seven large photometric outbursts, beginning at $r_H \sim$ 4 au, punctuated the rising lightcurve, with amplitudes up to about 5 magnitudes and becoming more frequent as $r_H$ decreased. 

\item The coma developed a distinctive ``horned" morphology in outburst, caused by radiation pressure repulsion of dust ejected at $\sim$375 m s$^{-1}$ from the sunward hemisphere of the nucleus.   The optically dominant dust particles have radii $\gtrsim$10 $\mu$m.

\item Each outburst has a sawtooth shape (steep rise and slow decline), with a rise time of hours followed by steady decay over 10 to 20 days.  The dust particle size distribution inferred from the best characterized outburst (outburst ``D") is a differential power law with index 4.2$\pm$0.3.

\item The median  mass per outburst in 12P, $M = 2\times10^9$ kg, is $\sim$10$^5$ times the typical outburst mass in well-studied comet 67P/Churyumov-Gerasimenko and, indeed, rivals the 
entire mass lost from 67P over the course of an orbit.

\item The kinetic energy per unit outburst mass is comparable to, or less than, the  ($\sim10^5$ J kg$^{-1}$) latent heat released by the crystallization of ice, consistent with crystallization as the driving process.  The outburst power is $P \sim 10^{10}$ W, too large to be sustained by equilibrium sublimation from the nucleus.
\end{itemize}

\clearpage

\appendix{Appendix A: Photometry Model}
\label{Appendix1}
\setcounter{equation}{0}
\renewcommand{\theequation}{A\arabic{equation}}

\begin{figure}
\epsscale{0.75}
\plotone{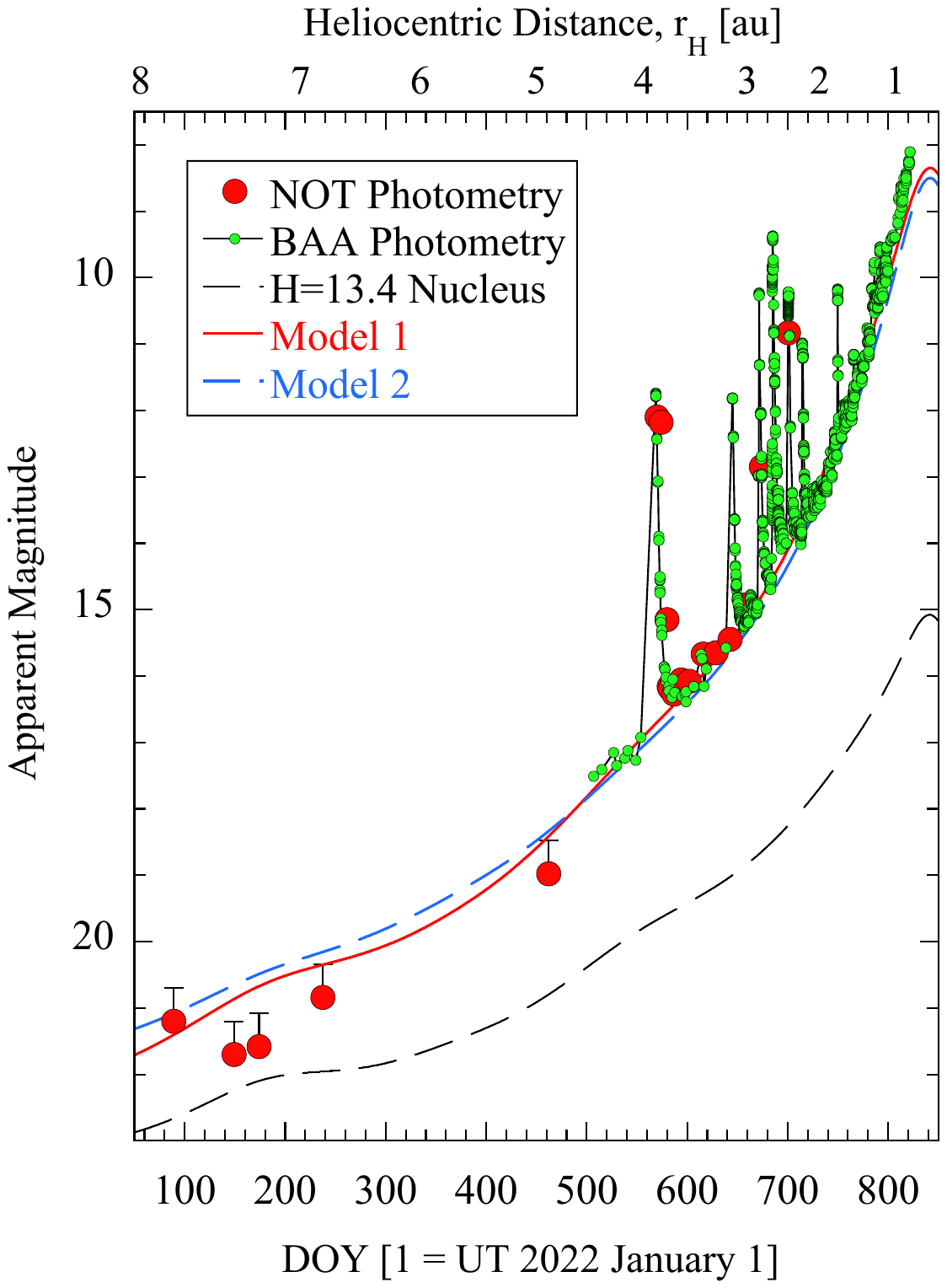}
\caption{Composite of BAA (small green circles) and NOT (large red circles) photometry  as a function of time expressed as Day of Year.  All measurements refer to a 9\arcsec~photometry aperture except those shown with upward-pointing error bars, which denote limits to the brightness forced by the use of smaller photometry apertures.  The dashed black line shows the expected variation of a bare nucleus having absolute magnitude $H$ = 13.4.  Model 1 (solid red line) shows a fit with heliocentric and spatial indices $m$ = 4.0, $n$ = 2, respectively, while Model 2 (dashed blue line) has $m$ = 4.5, $n$ = 1 (c.f., Equation \ref{R_c}).  Both models assume phase function 0.02 magnitudes per degree.     \label{NOT_BAA_4}}
\end{figure}

\begin{figure}
\epsscale{0.99}
\plotone{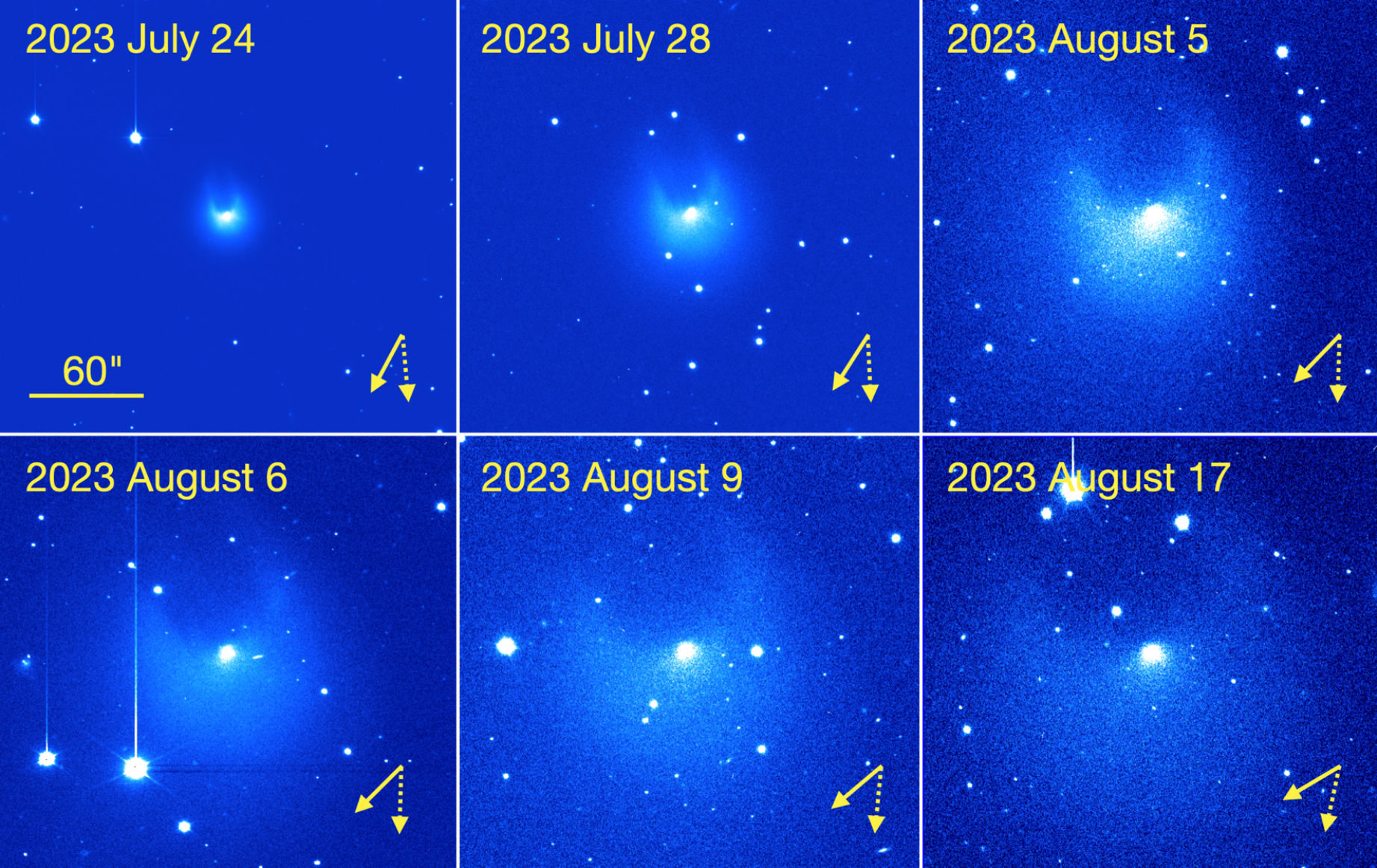}
\caption{R filter images of 12P Outburst A, each of 90 s integration from the NOT. Each panel has North to the top and East to the left.  A 60\arcsec~scale bar, corresponding to about 150,000 km at the comet, applies to all panels and is shown in the upper left.  The  projected antisolar direction and the negative of the heliocentric velocity vector are shown as solid and dotted arrows, respectively.  \label{horns}}
\end{figure}

\begin{figure}
\epsscale{0.9}
\plotone{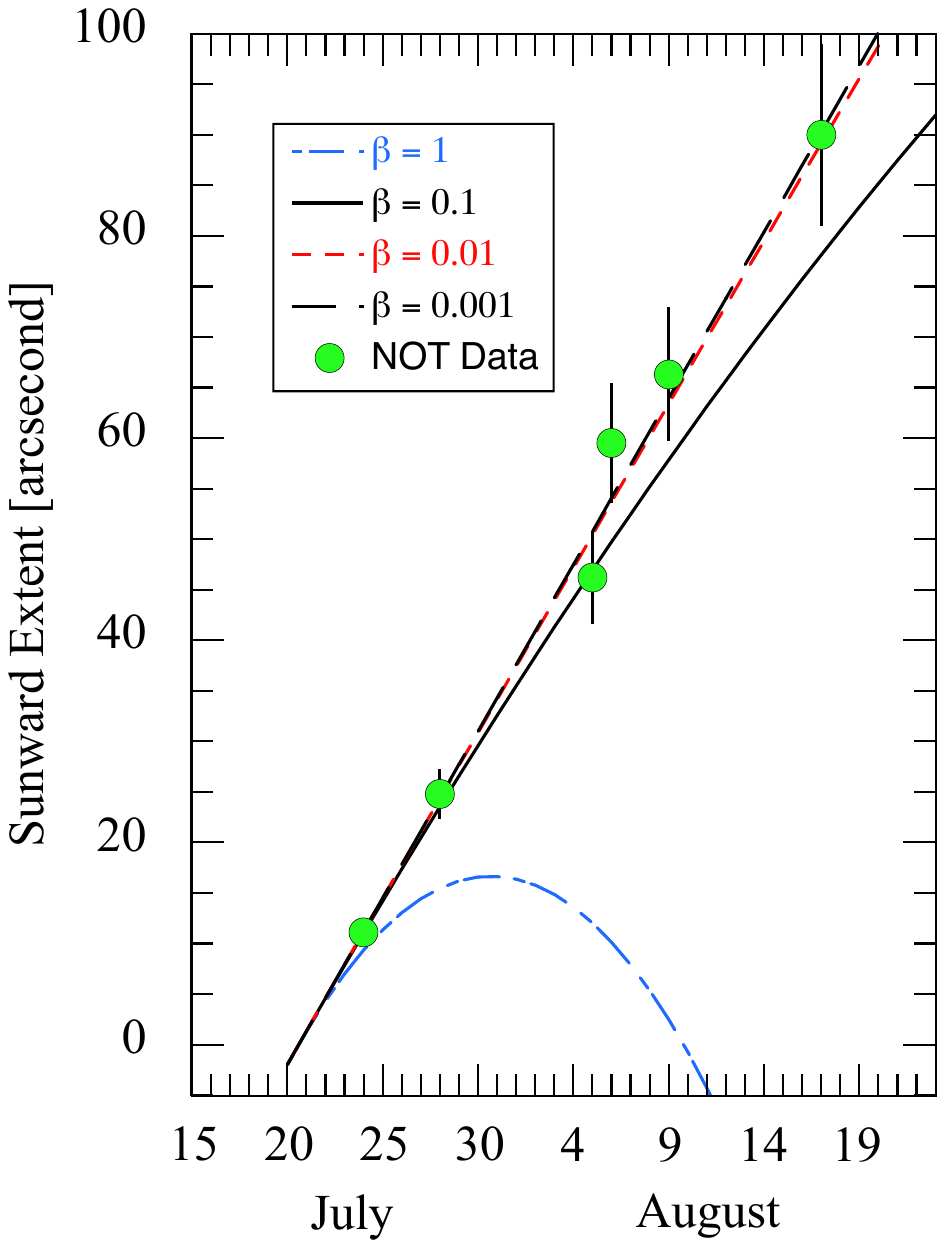}
\caption{Sunward extent of the coma in the 2023 July - August period as a function of date.  Lines in the figure show simple radiation pressure models for particles having radiation pressure efficiencies $\beta = 10^{-3}, 10^{-2}, 10^{-1}$ and 1, as marked.    \label{nose_length}}
\end{figure}

\begin{figure}
\epsscale{0.85}
\plotone{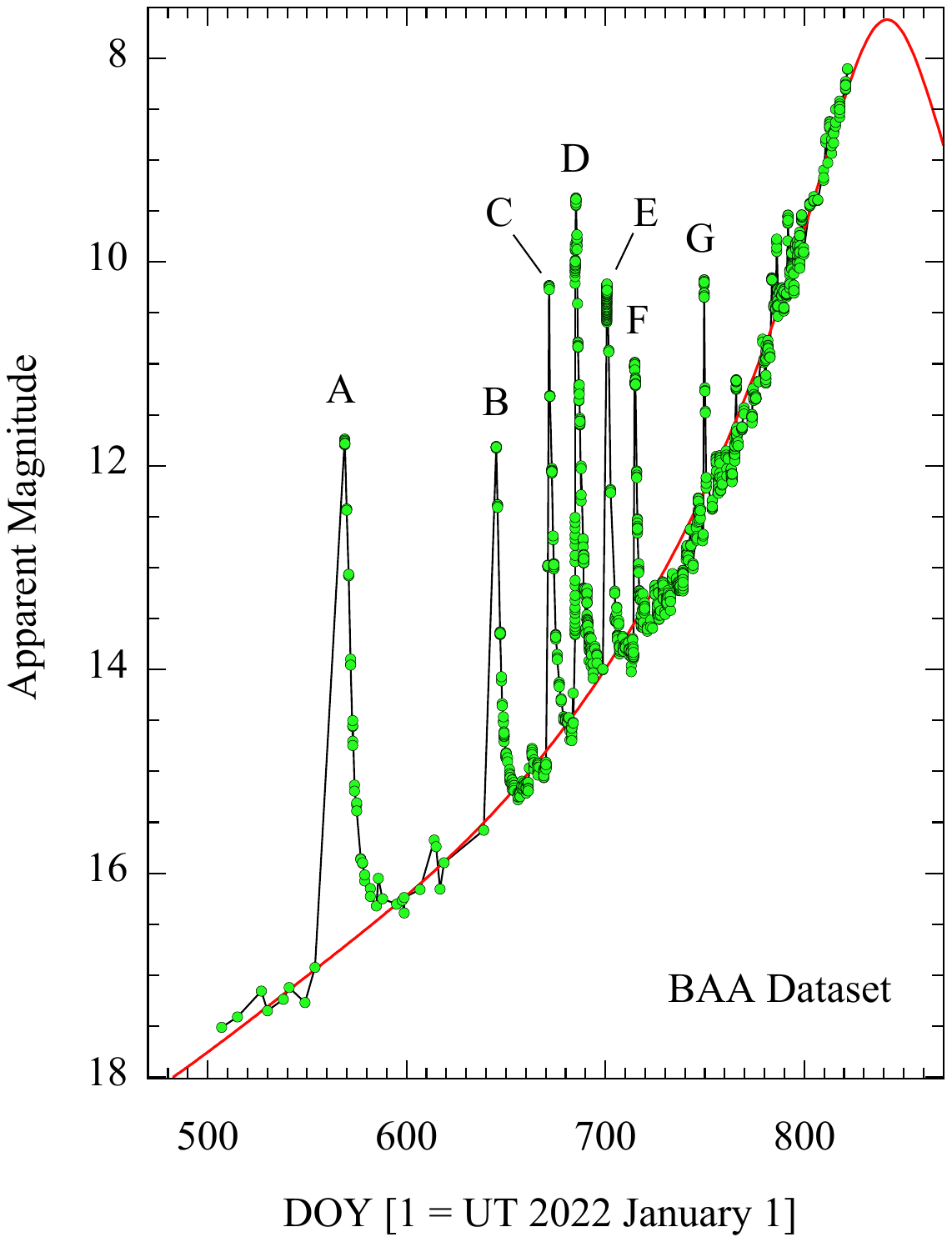}
\caption{Outburst photometry from BAA as a function of time expressed as Day of Year.  The major outbursts are labeled A - G.  A fit of Equation \ref{sublimation} to the non-outburst photometry with $m$ = 4.0, $n$ = 2, is shown as a solid red line.   \label{BAA}}
\end{figure}

\begin{figure}
\epsscale{0.9}
\plotone{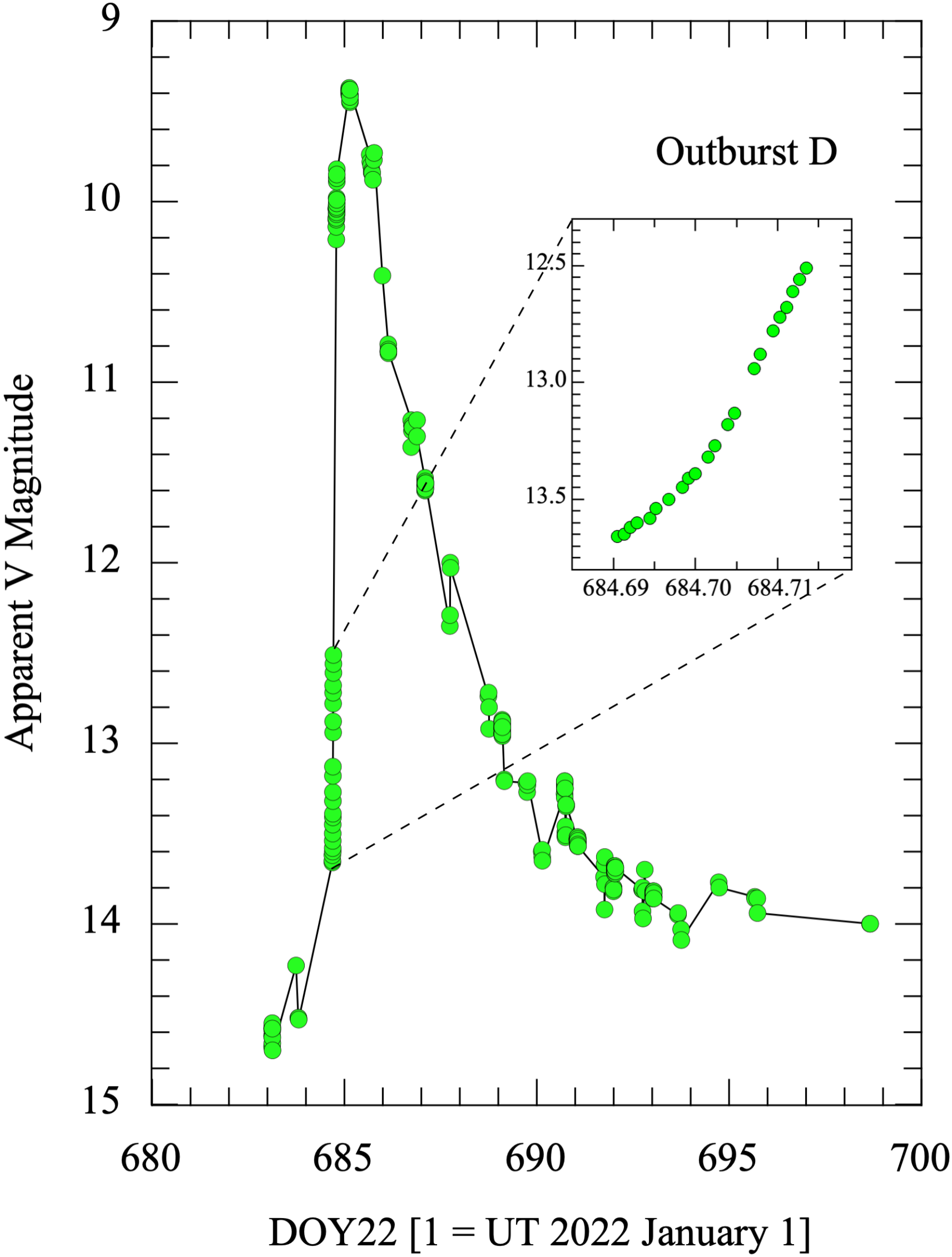}
\caption{Close-up of Outburst D (c.f., Figure \ref{BAA}) with a black line added to guide the eye. The zoom box shows the ultra rapid rise portion of the lightcurve with tick mark separation on the x-axis of 0.005 days or about 7 minutes. \label{outburst_D}}
\end{figure}

\begin{figure}
\epsscale{0.9}
\plotone{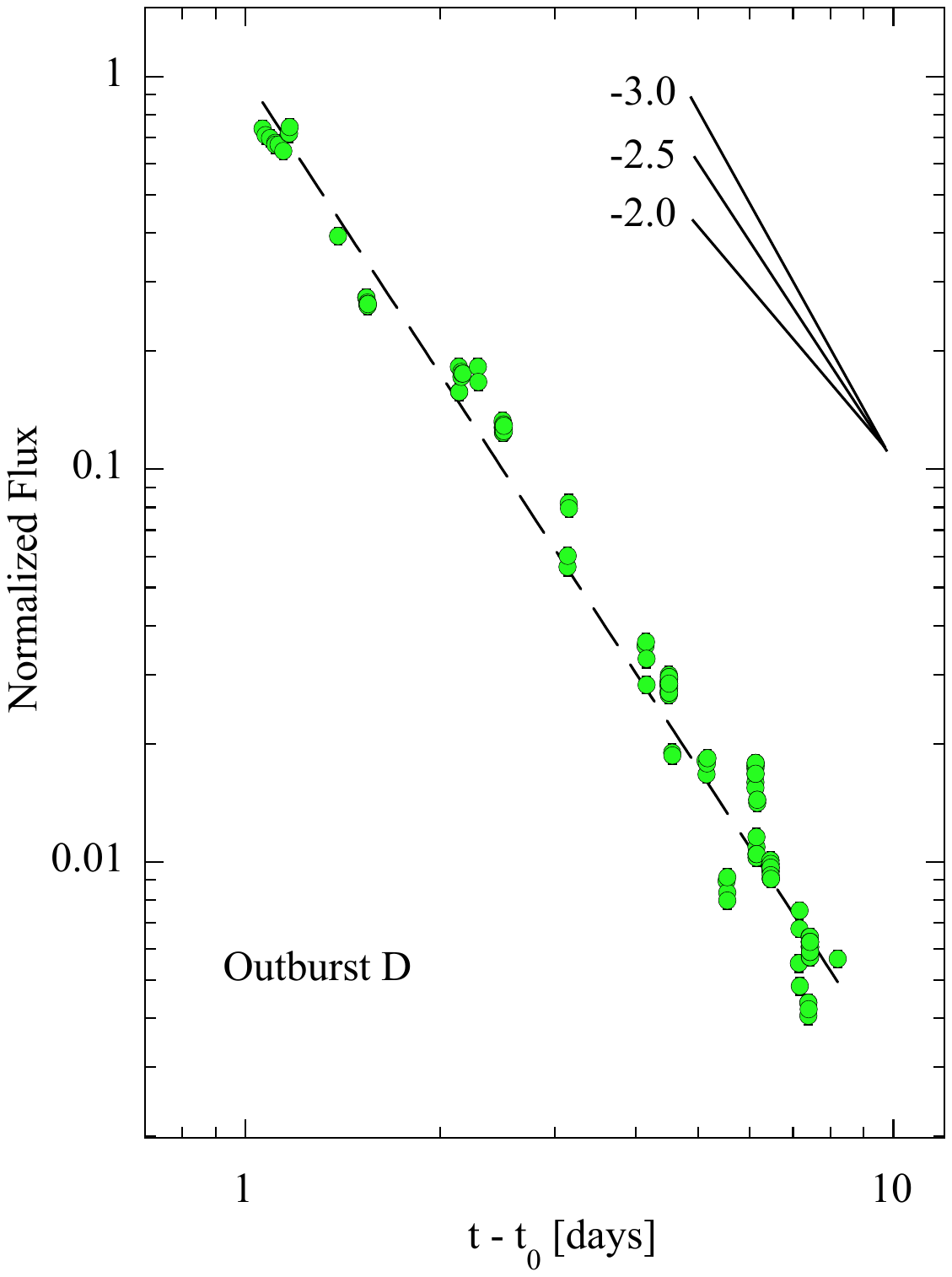}
\caption{Fading portion of the Outburst D lightcurve with background subtracted, converted to flux normalized at the peak, as a function of time measured from the peak $t_0$ = 685.1.  The dashed line shows a weighted least-squares power-law fit of slope -2.5, giving differential size index $\gamma$ = 4.2$\pm$0.2.   \label{Burst_D}}
\end{figure}

\begin{figure}
\epsscale{0.9}
\plotone{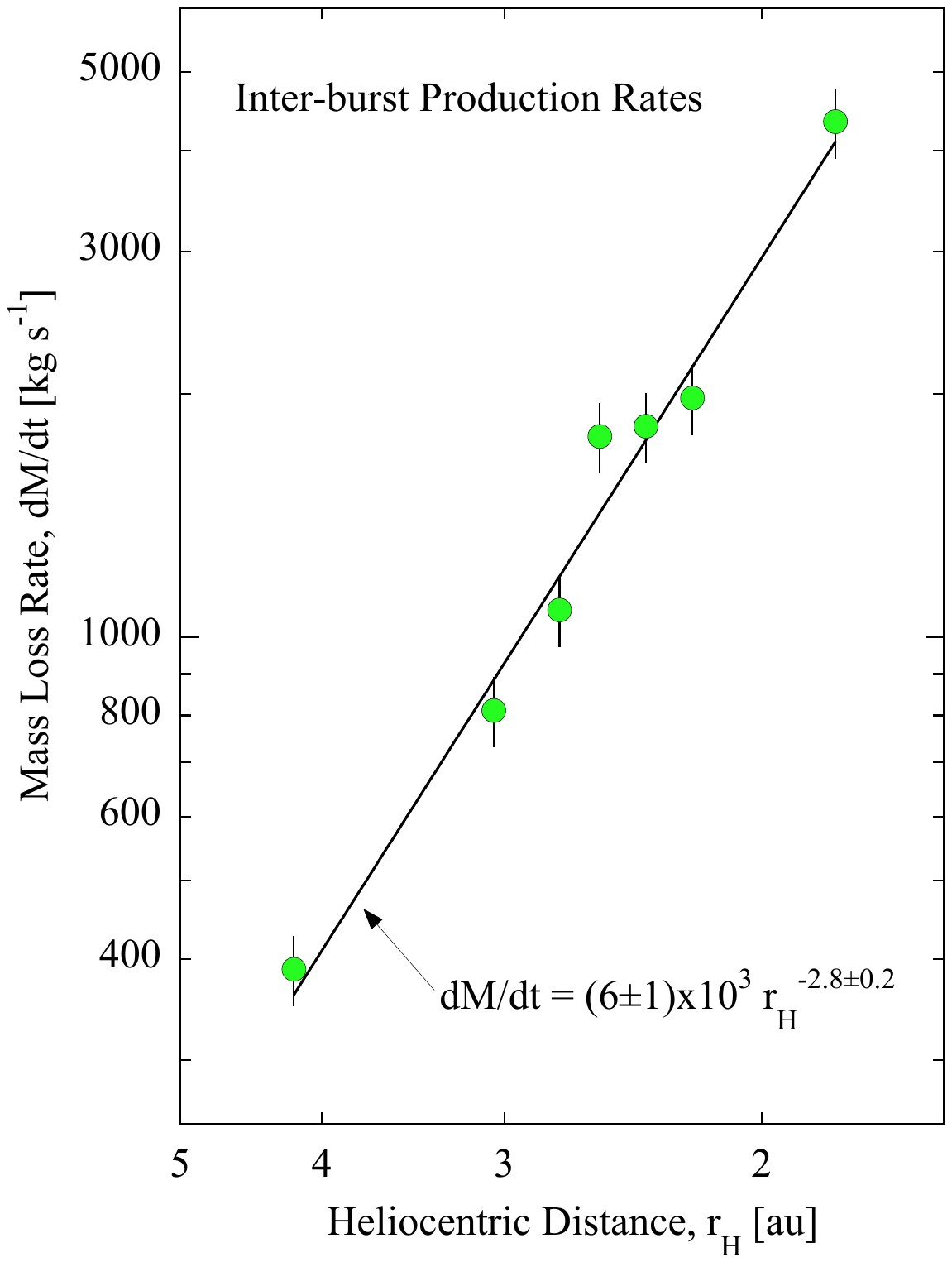}
\caption{Mass loss rate between outbursts as a function of heliocentric distance.  The solid black line shows a least squares power-law fit to the data. \label{dMbdt_vs_rH}}
\end{figure}

\begin{figure}
\epsscale{0.95}
\plotone{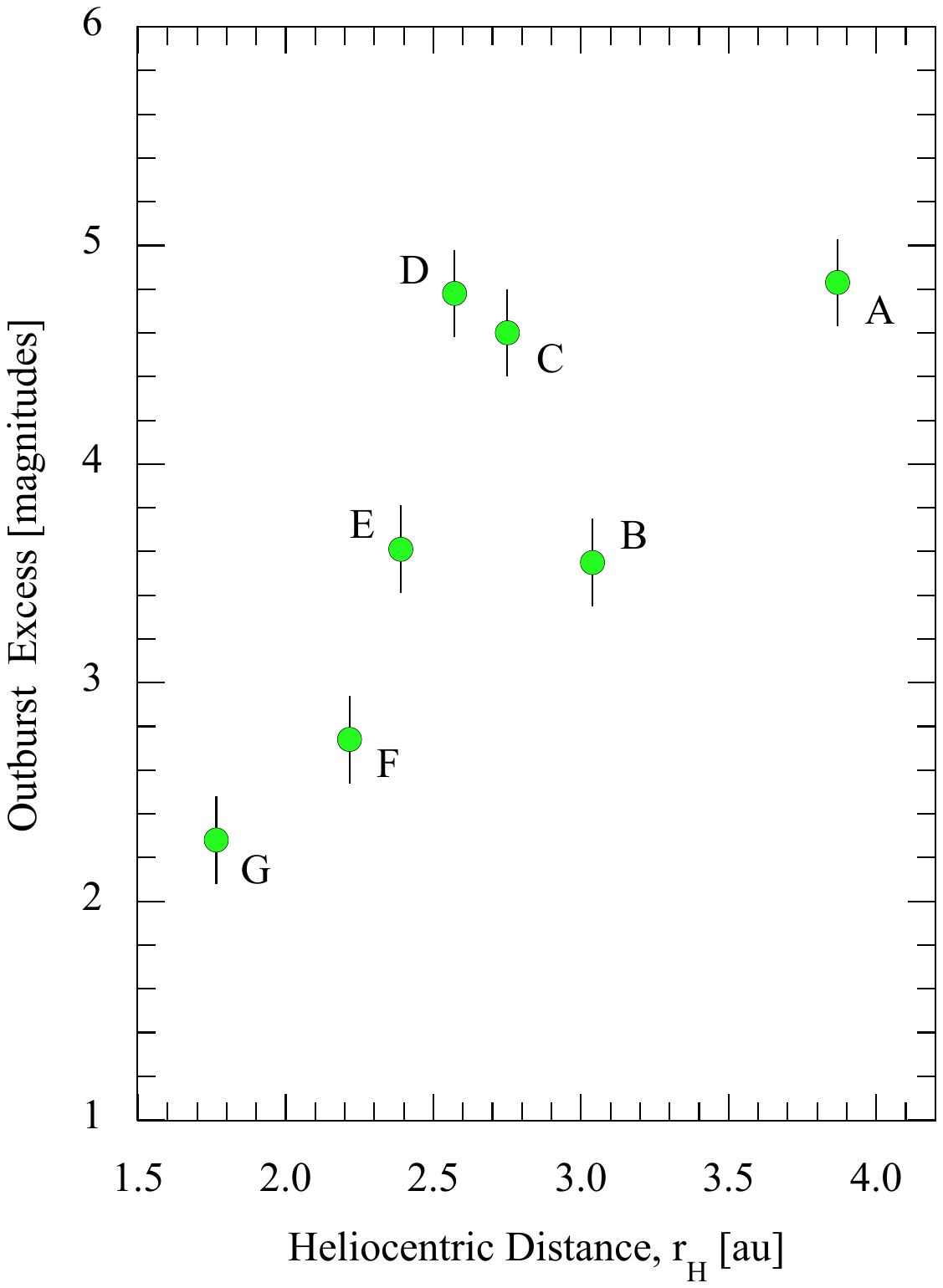}
\caption{The excess brightness of each outburst is plotted as a function of the heliocentric distance at the outburst time, with $\pm$0.2 magnitudes (1 $\sigma$) uncertainties on the excess.  Letters mark the individual outbursts.   \label{excess}}
\end{figure}

The apparent magnitude provides a measure of the sum of the scattering cross-sections of all the solid particles falling within the projected photometry aperture, which we denote $C$ [m$^2$].  The relation is given by

\begin{equation}
p_{R} \Phi(\alpha) C = 2.25\times10^{22} \pi r_H^2 \Delta^n 10^{-0.4(R_c - R_{\odot})}
\label{inversesquare}
\end{equation}

\noindent  in which $p_{R}$ is the geometric albedo at the wavelength of observation, $\Phi(\alpha)$ is the phase function at phase angle $\alpha$, $r_H$ and $\Delta$ are the heliocentric and geocentric distances expressed in astronomical units, and $R_c$ and $R_{\odot}$ = -27.12  \citep{Wil18} are the magnitudes of the comet and the Sun, respectively.  For a point source, or for any source subtending an angle smaller than the photometry aperture, the spatial index, $n$, takes on the value $n$ = 2, reflecting the standard inverse-square law dependence expected of an object having a fixed cross-section.  Index $n$ takes a value $n <$ 2 when the source occupies a solid angle larger than the photometry aperture.  For example,  the case of a steady-state outgassing gives a coma with $n$ = 1 \citep{Jew87}.  Comet 12P is evidently not in steady state, allowing the possibility that $n$ could exist outside the 1 $\le n \le$ 2 range. In any case, $n$ is clearly a function of time, with larger values corresponding to the more centrally condensed morphology in the early NOT data and smaller values corresponding to the outburst phases. We compute models with  $n$ = 1 and $n$ = 2, to show the sensitivity of the results to this parameter (Figure \ref{NOT_BAA_4}).

Neither the albedo nor the phase function of the dust in 12P have been measured.  Measurements of other comets suggest $p_V$ = 0.04 to 0.1 and a relatively flat backscattering phase function, with gradient $\beta \sim$ 0.02 to 0.04 magnitudes per degree over the 0 $\le \alpha \le$ 30\degr~phase angle range \citep{Mee87}.  We here assume $p_{\lambda}$ = 0.04,  $\Phi(\alpha) = 10^{-0.4 \beta \alpha}$ with $\beta$ = 0.02 magnitude degree$^{-1}$ and note that the phase angles in the range 500 $\le DOY \le$ 700 vary over a modest range  (7\degr~$\le \alpha \le$ 22\degr; Table \ref{geometry}), so that phase corrections are small compared to the large brightness variations observed. Should the albedo $\times$ phase function product $p_{\lambda}\Phi(\alpha)$ one day be measured for 12P, the rescaling of the cross-sections can be simply made using Equation \ref{inversesquare}.

Since the quantity of physical interest is $C$, the scattering cross-section of the dust within the projected photometry aperture, it is useful to rewrite Equation \ref{inversesquare}  as

\begin{equation}
R_c(r_H) = R_{\odot} - 2.5\log_{10}\left(\frac{p_V \Phi(\alpha) C(r_H)}{2.25\times10^{22} \pi r_H^2 \Delta^n}\right).
\label{R_c}
\end{equation}

Equation \ref{R_c}  provides a simple model of the comet lightcurve, from which the apparent magnitude, $R_c(r_H)$, can be calculated if the distance dependent cross section, $C(r_H)$, is known.

\appendix{Appendix B: Sublimation Model}
\setcounter{equation}{0}
\renewcommand{\theequation}{B\arabic{equation}}

We calculate simple photometric models assuming that the nucleus sublimates in equilibrium with sunlight at each heliocentric distance.  The objective is to understand the general brightening trend in 12P in terms of surface processes.  Neglecting the conduction of energy  into the nucleus, the power absorbed from the Sun is distributed between radiant heat and bond-breaking energy needed to sublimate the ice.  We write

\begin{equation}
\frac{L_{\odot}}{4\pi r_H^2}(1-A) \cos(\theta) = \varepsilon \sigma T^4 + L(T) f_s(T)  
\label{sublimation}
\end{equation}

\noindent in which $L_{\odot} = 4\times10^{26}$ W is the solar luminosity, $r_H$ is the heliocentric distance expressed in meters, $A$ is the Bond albedo, $\varepsilon$ is the emissivity of the nucleus, $\sigma = 5.67\times10^{-8}$ W m$^{-2}$ K$^{-4}$ is the Stefan-Boltzmann constant, $T$ is the nucleus temperature, $f_S$ [kg m$^{-2}$ s$^{-1}$] is the rate of sublimation per unit area of surface and $L(T)$ is the latent heat of sublimation. Quantity $\theta$ is  the angle between the direction to the Sun and the local normal to the surface.  For simplicity we assume $A$ = 0, $\varepsilon$ = 1.  The equation is solved iteratively using an additional relation for the temperature dependence of $f_s(T)$ based on the Clausius-Clapeyron equation for the slope of the solid-gas phase boundary.

Strictly, Equation \ref{sublimation} should be solved for each location on the nucleus and the resulting model mass loss rate obtained by integration over the whole  surface.  In reality, this is impossible, because we do not know the shape and size of the nucleus. Instead, we note the following simple approximations for the average value of $\cos(\theta)$.   The minimum  temperature and equilibrium mass loss rate are expected for a spherical, isothermal nucleus, where $\overline{\cos(\theta)}$ = 1/4.  The maximum temperature and specific mass loss rate occur at the sub-solar point, where  $\overline{\cos(\theta)}$ = 1. A spherical nucleus with a Sun-heated dayside and zero temperature nightside would have $\overline{\cos(\theta)}$ = 1/2. While real nuclei are irregularly shaped, not spherical, the observation that mass loss from 12P and most comets is concentrated on the Sun-facing side of the nucleus suggests that the latter is the best approximation for solution of Equation \ref{sublimation}. 


\acknowledgments
We thank Nick James and the BAA Comet Section for providing data in digital form, Pedro Lacerda for reading the manuscript and the anonymous referee for a helpful review.





\begin{deluxetable}{lcccrrrrrll}
\tabletypesize{\scriptsize}
\tablecaption{Observing Geometry 
\label{geometry}}
\tablewidth{0pt}
\tablehead{\colhead{UT Date} & \colhead{DOY\tablenotemark{a}} & \colhead{$\nu$\tablenotemark{b}}  & \colhead{$r_H$\tablenotemark{c}} & \colhead{$\Delta$\tablenotemark{d}}  & \colhead{$\alpha$\tablenotemark{e}} & \colhead{$\theta_{-\odot}$\tablenotemark{f}}  & \colhead{$\theta_{-V}$\tablenotemark{g}} & \colhead{$\delta_i$\tablenotemark{h}}  & \colhead{Note}  }

\startdata
2022 Mar 23.2 	&	82.2	&	212.2	&	7.939	&	8.030	&	7.1	&	273.2	& 220.6	&	6.6 & Not detected\\
2022 Mar 30.2 	&	89.2	&	212.3	&	7.890	&	7.903	&	7.2	&	268.2	&	220.0	&	6.5	& \\

2022 May 29.1 	&	149.1	&	213.6	&	7.456	&	6.930	&	6.9	&	214.7	&	207.1	&	1.3	& \\
2022 Jun 23.1	&	174.1 &	214.1	&	7.271	&	6.671	&	6.8	&	183.6	&	198.6	&	-2.1	&  \\
2022 Jul 30.0   & 210.0 & 215.0 & 6.991 & 6.496 & 7.5 & 137.6 & 185.9 & -6.6& \\
2022 Aug 25.6  	&	237.6	&	215.8	&	6.783	&	6.499	&	8.4	&	110.8	&	179.0	&	-8.3 &	 \\
2023 Apr 7.2	&	462.2 &	223.8	&	4.899	&	4.975	&	11.6	&	269.4 &	229.2	&	9.7	&  On star \\
2023 Apr 25.0	&	480.0 &	224.8	&	4.732	&	4.669	&	12.3	&	255.1	&	226.5	&	8.0	& On star\\
2023 Jul 24.0 	&	570.0	&	230.8	&	3.849	&	3.538	&	15.1	&	151.9 &	185.6	&	-11.1	&  \\
2023 Jul 28.0	&	574.0	&	231.1	&	3.808	&	3.507	&	15.3	&	146.6	&	183.0	&	-11.9	&  \\
2023 Aug 1.9	&	579.9	&	231.5	&	3.756	&	3.470	&	15.5	&	140.0	&	179.9	&	-12.9	&  \\
2023 Aug 5.1	&	582.1	&	231.8	&	3.724	&	3.450	&	15.7	&	136.1	&	178.1	&	-13.5	&  \\
2023 Aug 5.9    & 582.9	&	231.9	&	3.714	&	3.443	&	15.7	&	134.9	&	177.6	&	-13.6	&  \\
2023 Aug 9.9    & 586.9 & 232.3 & 3.671 & 3.415 & 15.9 & 129.8 & 175.2 & -14.3 & \\
2023 Aug 17.1   & 594.1 & 232.9 & 3.597 & 3.370 & 16.3 & 121.3 & 171.5 & -15.3 & \\
2023 Aug 25.1   & 602.1 & 233.7 & 3.512 & 3.322 & 16.7 & 112.0 & 167.7 & -16.3 & \\
2023 Sep 7.9    & 615.9 & 235.0 & 3.371 & 3.246 & 17.4 & 98.1 & 162.7 & -17.2 & \\
2023 Sep 20.9   & 628.9 & 236.6 & 3.217 & 3.164 & 18.1 & 84.6 & 159.1 & -17.3 & \\
2023 Oct 4.8    & 642.8 & 238.3 & 3.060 & 3.078 & 18.8 & 72.3 & 157.6 & -16.3 & \\
2023 Oct 19.8   & 657.8 & 240.4 & 2.889 & 2.974 & 19.5 & 60.2 & 158.8 & -14.2 & \\
2023 Nov 4.8    & 673.8 & 242.9 & 2.702 & 2.847 & 20.4 & 48.2 & 164.6 & -10.7 & \\
2023 Dec 1.8    & 700.8 & 247.8 & 2.388 & 2.601 & 22.3 & 30.5 & 192.4 & -2.7  & \\

\enddata


\tablenotetext{a}{Day of Year, 1 = UT 2022 January 1}
\tablenotetext{b}{True anomaly, in degrees}
\tablenotetext{c}{Heliocentric distance, in au}
\tablenotetext{d}{Geocentric distance, in au }
\tablenotetext{e}{Phase angle, in degrees }
\tablenotetext{f}{Position angle of projected anti-solar direction, in degrees }
\tablenotetext{g}{Position angle of negative heliocentric velocity vector, in degrees}
\tablenotetext{h}{Angle from orbital plane, in degrees}
\end{deluxetable}

\clearpage

\begin{deluxetable}{llccrrrrrr}
\tabletypesize{\scriptsize}
\tablecaption{Major Outbursts 
\label{outbursts}}
\tablewidth{0pt}
\tablehead{\colhead{ID\tablenotemark{a}} & UT Date\tablenotemark{b} & \colhead{DOY\tablenotemark{c}} & \colhead{$V_{peak}$\tablenotemark{d}} & \colhead{$V_{base}$\tablenotemark{e}} & \colhead{$\Delta$V\tablenotemark{f}}  & \colhead{$C_{peak}$\tablenotemark{g}}& \colhead{$C_{base}$\tablenotemark{h}}& \colhead{$\Delta$C\tablenotemark{i}} & \colhead{Mass\tablenotemark{j}}   }

\startdata
A & 2023 Jul 21    & 567 & 11.74 & 16.57 & 4.83 & 1.10$\times10^{11}$ & 1.28$\times10^{9}$ & 1.08$\times10^{11}$ & 2.2$\times10^{9}$\\
B & 2023 Oct 06    & 644 & 11.81 & 15.36 & 3.55 & 5.66$\times10^{10}$ & 2.15$\times10^9$ & 5.44$\times10^{10}$ & 1.1$\times10^{9}$\\
C & 2023 Nov 01    & 670 & 10.23 & 14.83 & 4.60 & 1.88$\times10^{11}$ & 2.71$\times10^9$ & 1.85$\times10^{11}$ & 3.7$\times10^{9}$\\
D & 2023 Nov 15    & 684 & 9.37  & 14.15 & 4.78 & 3.47$\times10^{11}$ & 4.25$\times10^9$ & 3.42$\times10^{11}$ & 6.8$\times10^{9}$  \\
E & 2023 Nov 30    & 699 & 10.22 & 13.83 & 3.61 & 1.30$\times10^{11}$ & 4.68$\times10^9$ & 1.25$\times10^{11}$ & 2.5$\times10^{9}$ \\
F & 2023 Dec 14    & 713 & 10.98 & 13.72 & 2.74 & 5.28$\times10^{10}$ & 4.23$\times10^9$ & 4.85$\times10^{10}$ & 1.0$\times10^{9}$ \\
G & 2024 Jan 18    & 748 & 10.17 & 12.45 & 2.28 & 6.21$\times10^{10}$ & 7.60$\times10^9$ & 5.45$\times10^{10}$ & 1.1$\times10^{9}$\\
\enddata

\tablenotetext{a}{Outburst label (see Figure \ref{BAA})}
\tablenotetext{b}{Estimated date of peak brightness}
\tablenotetext{c}{Day of Year of peak brightness}
\tablenotetext{d}{Peak magnitude}
\tablenotetext{e}{Base magnitude}
\tablenotetext{f}{Outburst range, magnitudes}
\tablenotetext{g}{Cross-section at outburst peak, m$^2$}
\tablenotetext{h}{Cross-section at outburst base, m$^2$}
\tablenotetext{i}{Change in cross-section due to outburst, m$^2$}
\tablenotetext{j}{Mass (kg), from Equation \ref{mass}}

\end{deluxetable}

\clearpage

%
%

\clearpage
%

%
%

\clearpage
\clearpage

\clearpage
\clearpage

\clearpage


\clearpage

\begin{thebibliography}{}

\bibitem[Bailey \& Emel'Yanenko(1996)]{Bai96} Bailey, M.~E. \& Emel'Yanenko, V.~V.\ 1996, \mnras, 278, 1087. doi:10.1093/mnras/278.4.1087

\bibitem[Bertaux(2015)]{Ber15} Bertaux, J.-L.\ 2015, \aap, 583, A38. doi:10.1051/0004-6361/201525992

\bibitem[Bobrovnikoff(1932)]{Bob32} Bobrovnikoff, N.~T.\ 1932, \pasp,  44, 261, 296. doi:10.1086/124249

\bibitem[Davidsson et al.(2022)]{Dav22} Davidsson, B.~J.~R., Samarasinha, N.~H., Farnocchia, D., et al.\ 2022, \mnras, 509, 3065. doi:10.1093/mnras/stab3191
67P/

\bibitem[Dones et al.(2015)]{Don15} Dones, L., Brasser, R., Kaib, N., et al.\ 2015, \ssr, 197, 191. doi:10.1007/s11214-015-0223-2

\bibitem[Emel'yanenko et al.(2013)]{Eme13} Emel'yanenko, V.~V., Asher, D.~J., \& Bailey, M.~E.\ 2013, Earth Moon and Planets, 110, 105. doi:10.1007/s11038-012-9413-z

\bibitem[Evans et al.(2018)]{Eva18} Evans, D.~W., Riello, M., De Angeli, F., et al.\ 2018, \aap, 616, A4. doi:10.1051/0004-6361/201832756


\bibitem[Fern{\'a}ndez et al.(2016)]{Fer16} Fern{\'a}ndez, J.~A., Gallardo, T., \& Young, J.~D.\ 2016, \mnras, 461, 3075. doi:10.1093/mnras/stw1532

\bibitem[Gritsevich et al.(2025)]{Gri25} Gritsevich, M., Weso{\l}owski, M., \& Castro-Tirado, A.~J.\ 2025, \mnras. doi:10.1093/mnras/staf219

\bibitem[Gudipati et al.(2023)]{Gud23} Gudipati, M.~S., Fleury, B., Wagner, R., et al.\ 2023, Faraday Discussions, 245, 467. doi:10.1039/D3FD00048F

\bibitem[Ishiguro et al.(2016)]{Ish16} Ishiguro, M., Kuroda, D., Hanayama, H., et al.\ 2016, \aj, 152, 169. doi:10.3847/0004-6256/152/6/169

\bibitem[Jewitt \& Meech(1987)]{Jew87} Jewitt, D.~C. \& Meech, K.~J.\ 1987, \apj, 317, 992. doi:10.1086/165347

\bibitem[Jewitt et al.(2016)]{Jew16} Jewitt, D., Mutchler, M., Weaver, H., et al.\ 2016, \apjl, 829, L8. doi:10.3847/2041-8205/829/1/L8

\bibitem[Jewitt et al.(2017)]{Jew17} Jewitt, D., Agarwal, J., Li, J., et al.\ 2017, \aj, 153, 223. doi:10.3847/1538-3881/aa6a57

\bibitem[Jewitt \& Kim(2020)]{Jew20} Jewitt, D. \& Kim, Y.\ 2020, PSJ, 1, 77. doi:10.3847/PSJ/abbef6

\bibitem[Jewitt(2021a)]{Jew21} Jewitt, D.\ 2021, \aj, 161, 261. doi:10.3847/1538-3881/abf09c

\bibitem[Jewitt et al.(2021b)]{Jew21b} Jewitt, D., Kim, Y., Mutchler, M., et al.\ 2021, \aj, Cometary Activity Begins at Kuiper Belt Distances: Evidence from C/2017 K2, 161, 4, 188. doi:10.3847/1538-3881/abe4cf





\bibitem[Kelley et al.(2021)]{Kel21} Kelley, M.~S.~P., Farnham, T.~L., Li, J.-Y., et al.\ 2021, PSJ, 2, 4, 131. doi:10.3847/PSJ/abfe11

\bibitem[Kronk(2003)]{Kro03} Kronk, G.~W.\ 2003, Cometography, by Gary W. Kronk,  ISBN 0521585058. Cambridge, UK: Cambridge University Press, p.852

\bibitem[Landolt(1992)]{Lan92} Landolt, A.~U.\ 1992, \aj, 104, 340. doi:10.1086/116242


\bibitem[Li et al.(2011)]{Li11} Li, J., Jewitt, D., Clover, J.~M., et al.\ 2011, \apj, 728, 31. doi:10.1088/0004-637X/728/1/31

\bibitem[Lin et al.(2017)]{Lin17} Lin, Z.-Y., Knollenberg, J., Vincent, J.-B., et al.\ 2017, \mnras, 469, S731. doi:10.1093/mnras/stx2768

\bibitem[Meech \& Jewitt(1987)]{Mee87} Meech, K.~J. \& Jewitt, D.~C.\ 1987, \aap, 187, 585

\bibitem[Meyer et al.(2020)]{Mey20} Meyer, M., Kobayashi, T., Nakano, S., et al.\ 2020, arXiv:2012.15583. doi:10.48550/arXiv.2012.15583


\bibitem[M{\"u}ller et al.(2024)]{Mul24} M{\"u}ller, D.~R., Altwegg, K., Berthelier, J.-J., et al.\ 2024, \mnras, 529, 2763. doi:10.1093/mnras/stae622

\bibitem[Pancino et al.(2022)]{Pan22} Pancino, E., Marrese, P.~M., Marinoni, S., et al.\ 2022, \aap, 664, A109. doi:10.1051/0004-6361/202243939

\bibitem[Prialnik \& Bar-Nun(1992)]{Pri92} Prialnik, D. \& Bar-Nun, A.\ 1992, \aap, 258, L9

\bibitem[Prialnik \& Jewitt(2024)]{Pri24} Prialnik, D., and Jewitt, D. \ 2024. In Comets III, editor K. Meech et al., University of Arizona Press, Tucson, Arizona, pp. 823-843

\bibitem[Schambeau et al.(2017)]{Sch17} Schambeau, C.~A., Fern{\'a}ndez, Y.~R., Samarasinha, N.~H., et al.\ 2017, \icarus,  284, 359. doi:10.1016/j.icarus.2016.11.026

\bibitem[Sekanina et al.(1992)]{Sek92} Sekanina, Z., Larson, S.~M., Hainaut, O., et al.\ 1992, \aap, 263, 367



\bibitem[Willmer(2018)]{Wil18} Willmer, C.\ 2018, \apjs, 236, 47. doi:10.3847/1538-4365/aabfdf

\bibitem[Ye et al.(2020)]{Ye20}Ye, Q. et al. 2020. Research Notes of the AAS,  4, id.101



\end{thebibliography}
\end{document}